%% file: megeath.tex
\documentclass[cup6a]{cupbook}
\usepackage{psfig}
\usepackage{epsf}
\usepackage{graphicx}

\usepackage{natbib}
\title[The Formation of Distributed and Clustered Stars in Molecular Clouds]
      {The Formation of Distributed and Clustered Stars in Molecular Clouds}
\author[S. T. Megeath]{S. T. Megeath, University of Toledo \and Zhi-Yun Li, University of Virginia  \and Aake Nordlund, Niels Bohr Institute}
\date{31 Oct 2009}

\newcommand\PSFIGG[2]{\centerline{\psfig{#1}\psfig{#2}}}

\begin{document}

\pagenumbering{roman}
\maketitle
\tableofcontents
\cleardoublepage
\pagenumbering{arabic}

\chapter[The Formation of Distributed and Clustered Stars in Molecular Clouds]%
        {The Formation of Distributed and Clustered Stars\\ in Molecular Clouds}

\author[S. T. Megeath]{University of Toledo \and Zhi-Yun Li, University of Virginia  \and Aake Nordlund, Neils Bohr Institute}

\section{Introduction}

During the last two decades, the focus of star formation research has
shifted from understanding the collapse of a single dense core into a
star to studying the formation hundreds to thousands of stars in
molecular clouds.  In this chapter, we overview recent observational
and theoretical progress toward understanding star formation on the
scale of molecular clouds and complexes, i.e the macrophysics of star
formation \citep{2007ARA&A..45..565M}.  We begin with an overview of
recent surveys of young stellar objects (YSOs) in molecular clouds and
embedded clusters, and we outline an emerging picture of cluster
formation.  We then discuss the role of turbulence to both support
clouds and create dense, gravitationally unstable structures, with an
emphasis on the role of magnetic fields (in the case of distributed
stars) and feedback (in the case of clusters) to slow turbulent decay
and mediate the rate and density of star formation.  The discussion is
followed by an overview of how gravity and turbulence may produce
observed scaling laws for the properties of molecular clouds, stars
and star clusters, and how the observed, low star formation rate may
result from self regulated star formation. We end with some concluding
remarks, including a number of questions to be addressed by future
observations and simulations.

\section{Observations of Clustered and Distributed Populations in Molecular Clouds}

Our knowledge of the distribution and kinematics of young stars,
protostars and dense cores in molecular clouds is being rapidly
improved by wide field observations at X-ray, optical, infrared, and
(sub)millimeter wavelengths
\citep{2007prpl.conf..361A,2007prpl.conf..313F}.  In this section we
present multiwavelength portraits of some of the most active star
forming regions near our Sun and discuss some implications of these
results.

\subsection{Molecular Cloud Surveys}

\begin{figure}[htp]
  \psfig{file=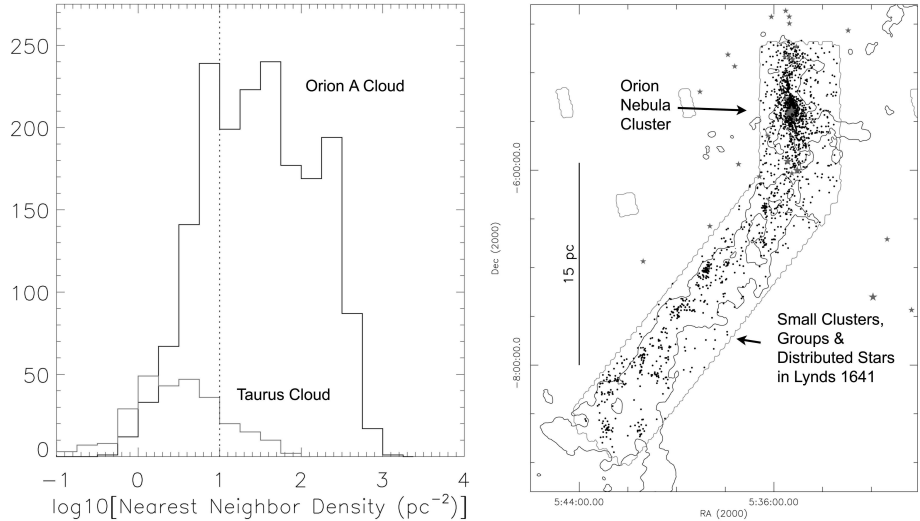,width=5 in}
 \caption{{\bf Left:} The distribution of nearest neighbor densities
  ($10/(\pi R_{10}^2)$, where $R_{10}$ is the distance to the 10th
  nearest neighbor) for the Orion A GMC and the Taurus dark cloud
  complex (K. Luhman, priv.\ comm.).  The dotted line gives the adopted
  threshold density for separating distributed stars from stars in
  groups and clusters.  {\bf Right:} The distribution of Spitzer
  identified young stellar objects with infrared excesses in the Orion
  molecular clouds. The gray line gives the outline of the survey
  fields, including the large hockey stick shaped field and three
  reference fields.  The black lines is the $A_V=3$ contour determined
  from an extinction map constructed from the 2MASS point source
  catalog (R. Gutermuth, priv.\ comm.).  The star symbols give the
  positions of OB stars from \citet{1994A&A...289..101B}. }
 \label{fig:oriona}
\end{figure}

The {\it Spitzer} space telescope, with its unparalleled ability to
detect IR-excesses from YSOs with disk and envelopes over wide fields,
is now producing the most complete censuses to date of young stellar
objects (YSOs) in nearby molecular clouds.  One example is the survey
of the Orion~A cloud, the most active star forming cloud within 450~pc
of the Sun (Megeath et al., in prep).  The observed distribution of
YSOs in this cloud exhibits structure on a range of spatial scales,
tracing both the overall filamentary morphology of the clouds and
smaller filamentary and clumpy sub-structures within the clouds, as
shown in Fig.~\ref{fig:oriona}.  This figure also shows a range of
observed stellar densities; YSOs can be found in large clusters containing
OB stars, in smaller groups and in relative isolation.

\begin{figure}[t]
\psfig{file=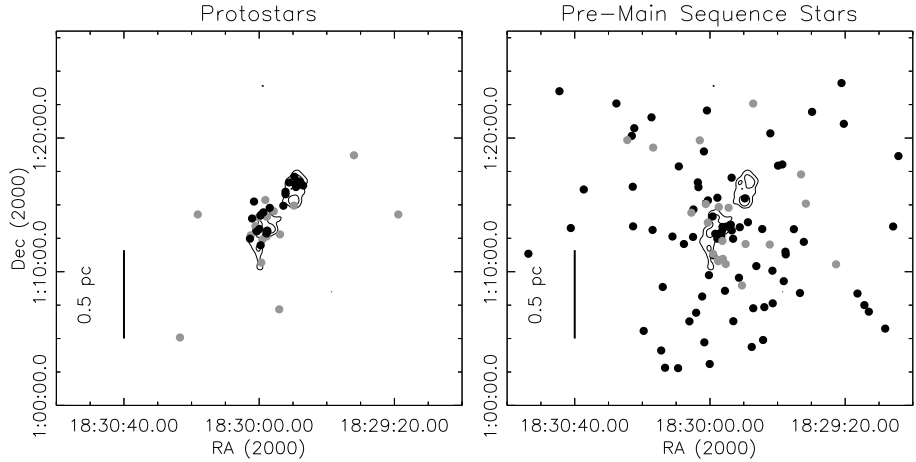,width=5. in}
\caption{The distribution of young stellar objects in the Serpens
 cloud core cluster at a distance of 260 pc
 \citep{2007ApJ...669..493W}.  The contours show the dense gas detected
 in an 850~$\mu$m SCUBA map \citep{2000ApJ...530L.115D}. The left panel
 shows the distribution of protostars separated into flat spectrum
 (gray markers) and Class~I/0 (black markers) sources. The right panel
 show the pre-main sequence stars with disks (black markers) and
 without disks (gray markers).  The pre-main sequence stars without
 disks were identified by their elevated X-ray emission
 \citep{2007A&A...463..275G}; see \citet{2007ApJ...669..493W} for a
 discussion of the completeness the displayed source selection. Note
 that the SCUBA map and Chandra images do not cover the entire
displayed region.}
\label{fig:serpens}
\end{figure}

\begin{figure}[htp]
 \psfig{file=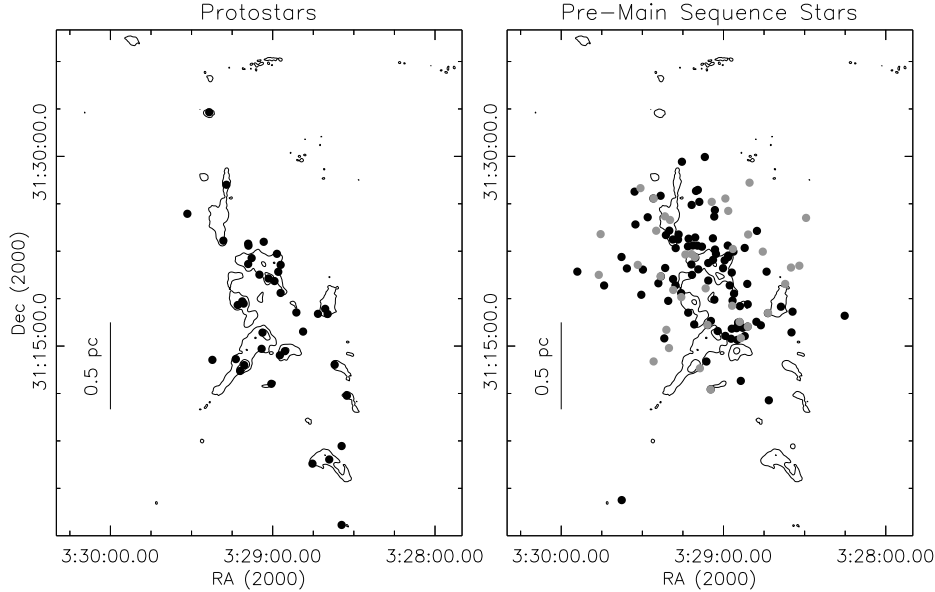,width=5. in}
 \caption{The distribution of YSOs in the NGC 1333 cluster at a
 distance of 250~pc. The contours show the dense gas traced in the
 850~$\mu$m SCUBA map \citep{2001ApJ...546L..49S}. The left panel
 shows the distribution of protostars
 (Gutermuth et al., 2007); the
 right panel shows the distribution of pre-main sequence stars with
 disks (black markers; Gutermuth et al., 2007);
 and diskless pre-main
 sequence stars with elevated X-ray emission \citep[gray
 markers;][]{2002ApJ...575..354G}.  Note that the Chandra survey of
 Getman et al.  does not cover the entire displayed field.}
 \label{fig:ngc1333}
\end{figure}
\nocite{2007arXiv0710.1860G}

\begin{figure}[htp]
 \psfig{file=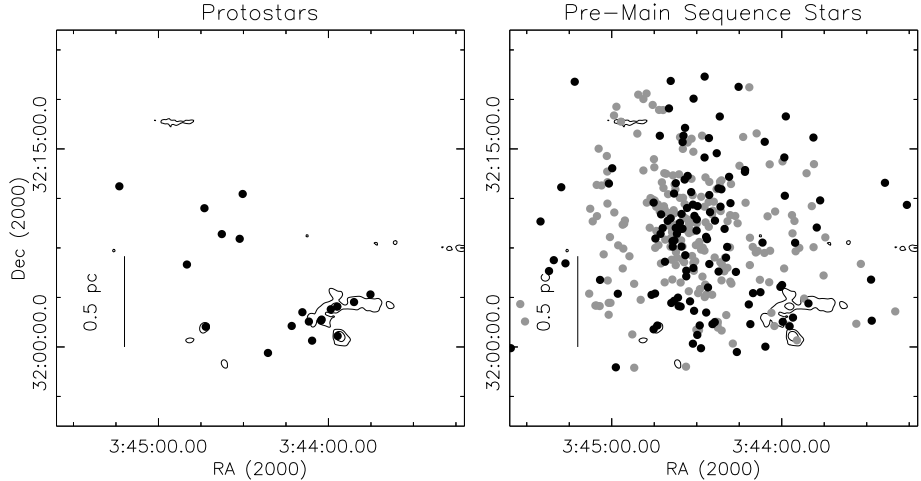,width=5. in}
 \caption{The distribution of YSOs in the IC 348 Cluster. The contours
show the dense gas as traced by the 850~$\mu$m SCUBA map
(J. DiFrancesco, priv.\ comm.).  The left panels show the distribution of
protostars (black markers); the right panel shows the distribution of
pre-main sequence stars with (black markers) and without (gray
markers) disks. The sample was taken from \citet{2007AJ....134..411M}.}
 \label{fig:ic348}
\end{figure}

\begin{figure}[htp]
 \psfig{file=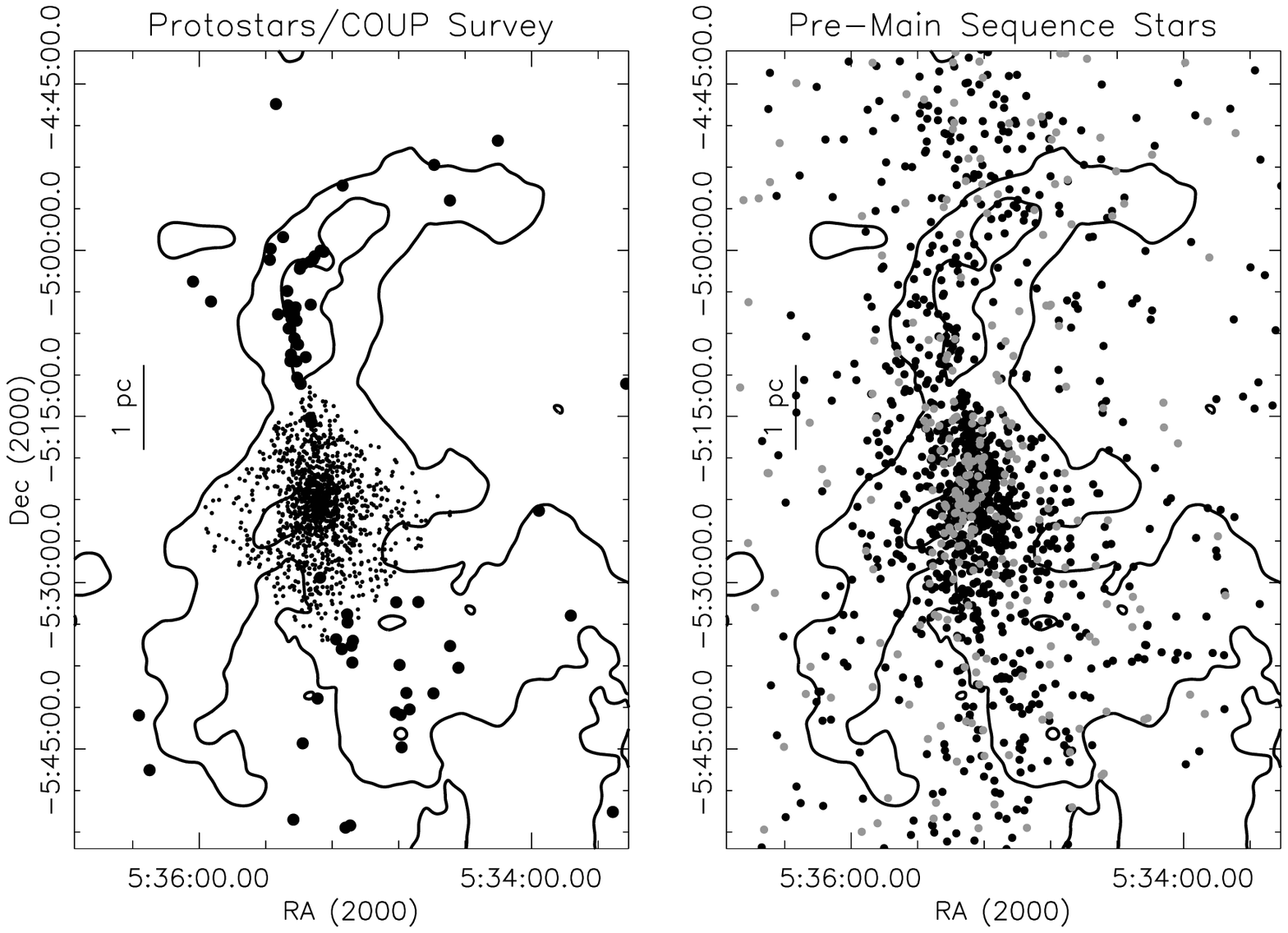,width=5. in}
 \caption{The distribution of YSOs in the Orion Nebula Cluster.  The
 contours show an extinction map constructed
 from the 2MASS PSC (R. Gutermuth, priv.\ comm.).  {\bf Left panel:}
 The large black circles are protostars identified in the Spitzer data
 (Megeath et al., in prep.);  no protostars are identified in the
 center of the Orion nebula where the 24~$\mu$m data is saturated.
 The small black dots are the detected X-ray sources from the Chandra
 Orion Ultradeep Project \citep{2005ApJS..160..379F}; this survey
 covered a diamond shaped region defined by the extent of the X-ray
 sources.  {\bf Right Panel} the distribution of Spitzer identified
 pre-main sequence stars with disks (black markers; Megeath et al. in
 prep) and pre-main sequence stars without disks (gray markers)
 identified by variability \citep{2001AJ....121.3160C}.}
 \label{fig:onc}
\end{figure}

Is there a preferred environment for star formation in giant molecular
clouds: do most low mass stars form in rare large clusters, the more
numerous small groups, or in a distributed population of more isolated
stars?  To some extent, this depends on the criteria used to
distinguish isolated and clustered stars; as shown in Orion~A, the
clusters are often not distinct objects but belong to an extended
distribution of stars (Fig.~\ref{fig:oriona}).  One approach is to
adopt a critical surface density for clustered stars, although there
is no clear motivation for a particular threshold density as
demonstrated by the continuous distribution of YSO surface densities
shown in Fig.~\ref{fig:oriona}.  In the Orion cloud complex (Orion A
and B), Megeath et al. (in prep) decomposed the observed distribution
of YSOs with IR-excesses into groups of 10 or more sources where each
member is in a contiguous region with a local surface density of 10
stars per pc$^{-2}$ or higher. They find that 44\% are in the 1000
member Orion Nebula Cluster (ONC), 18\% are in seven clusters with 30
to 100 YSOs and 9\% are in twelve groups of 30 to 10 YSOs.  They find
that 28\% of the YSOs with IR-excesses (639 objects in total) form a
distributed population outside the groups and clusters \citep[also
see][]{2007prpl.conf..361A}.  The number of YSOs in the ONC is
underestimated due to incompleteness, and consequently the actual
fraction of stars in the distributed population is somewhat lower
(Megeath et al. in prep).  Despite this bias, the surveys are showing
that even in giant molecular clouds (GMCs) forming massive stars, a
substantial number of YSOs are found in relative isolation; the
observed range of densities and environments must be explained by
models of cloud fragmentation and evolution.

\subsection{A Gallery of Embedded Clusters}

In Figs.~\ref{fig:serpens}, \ref{fig:ngc1333}, \ref{fig:ic348} \&
\ref{fig:onc}, we show the latest compilations of YSOs in four of the
nearest embedded (or partially embedded) clusters to the Sun.  The
most striking result is the diversity of configurations apparent in
the clusters.  The dense gas in the Serpens cluster is concentrated in
a 0.5~pc long filament divided into two main clumps; the protostars
are primarily found in the two clumps.  In contrast, the pre-main
sequence stars are more evenly distributed over a 1 pc diameter region
with only a modest concentration in the central region; this may be
the result of dynamical evolution due to gravitational interactions
between YSOs and due to the non-spherically symmetric cloud potential.
The dense gas in NGC~1333 forms a network of clumps and filaments
spread over a region more than~1 pc in diameter.  Both the protostars
and pre-main sequence stars follow this structure, with the protostars
concentrated in the dense gas and pre-main sequence stars found in the
immediate vicinity of the dense gas; this implies little dynamical
evolution in this region.  Unlike the previous two clusters, the IC348
cluster shows a circularly symmetric and centrally condensed
distribution. The molecular gas has been largely cleared from this
region; the protostars are concentrated in a filamentary gas structure
on the edge of the cluster.

The massive Orion Nebula Cluster (ONC) shows a particularly complex
morphology (Figs.~\ref{fig:oriona} \& Fig.~\ref{fig:onc}).  In the
center of the ONC is the densest known region of young stars in the
nearest 500~pc.  This central condensation is elongated and aligned
with the axis of the filamentary molecular cloud.  To the north of the
central condensation a filamentary distribution of protostars follows
the molecular cloud. It is surrounded by a more extended distribution
of pre-main sequence stars.  To the south, the molecular cloud appears
to have been partially swept up into a shell by the massive stars in
the center of the ONC. A more extended distribution of YSOs in this
regions fills the shell, indicating that these stars probably
originated in the expanding shell.

\subsection{The Distribution of Protostars in Embedded Clusters}

The distribution and spacing of protostars is an important constraint
on the physics of fragmentation and the potential for subsequent
interactions between protostars.  As shown previously, the protostars
trace the clumpy and filamentary distribution of the dense molecular
gas.  The protostars in embedded clusters are often closely packed,
with median projected nearest neighbor spacings ranging from 5000~AU
to 20000~AU \citep{2007ApJ...669..493W,2007AJ....134..411M}. In
a few cases, peaks are found in the distribution of projected
nearest neighbor distances for protostars
\citep{2007arXiv0710.1860G,2006ApJ...636L..45T}
and for dense cores
\citep{2006A&A...447..609S}.  These preferred separations
for the protostars are within a factor of three of the local Jeans length,
suggesting that the protostars may result from gravitational fragmentation
\nocite{2006ApJ...636L..45T} (Teixeira et al., 2006).
The detection of extremely dense groups of
protostars with $\sim 7$ objects in regions $\sim 10000$~AU diameter
suggest that hierarchical fragmentation operates in some cases
\citep{2007ApJ...669..493W,2007ApJ...667L.179T}.

Can the protostars in the observed dense groups interact?  In the
Serpens region the close spacing of protostars implies that the volumes
from which these protostars accrete are often densely packed, if not
overlapping \citep{2007ApJ...669..493W}.  In a study of a dense group
of protostars in Serpens, \citet{2007ApJ...669..493W} argued that the
high gas density \citep{2002A&A...392.1053O}, and the subvirial RMS
velocity of the protostars \citep[as measured in the BIMA N$_2$H$^+$
detections of the protostellar cores;][]{2000ApJ...537..891W} are
consistent with the protostars accreting a significant portion of
their mass through Bondi-Hoyle accretion.  Bondi-Hoyle accretion in
such a closely packed group may lead to the competitive accretion
predicted in some numerical models of turbulent clouds
\citep{2006MNRAS.370..488B}.  Furthermore, if such groups stay bound
for 400,000 years \citep[the estimated duration of the protostellar
phase][]{2007A&A...468.1009H}, each large (1000~AU) protostellar
envelope would typically experience one collision.  Further studies
are needed to find direct evidence for interactions and to ascertain
the fraction of stars that form in dense groups where the potential
for interactions exists.

\subsection{The Evolution of Embedded Clusters: An Observational Perspective}

A common trait of embedded clusters is a lack of circular symmetry:
they are often elongated and in many cases show substructure
\citep{2005ApJ...632..397G,2007prpl.conf..361A}.  This indicates that
clusters retain the imprint of the filamentary, clumpy structure of
the molecular clouds from which they formed; consequently they do not
appear to be relaxed, virialized clusters of stars.  This is supported
by observations of the kinematics of protostellar cores, which show
that their velocity dispersions are sub-virial \citep[from observation
of the NGC1333 and NGC2264
clusters;][]{2007ApJ...655..958W,2007A&A...464..983P}.  In the Orion
Nebula, \citet{2007arXiv0711.0391F} find that the velocities of the
pre-main sequence stars largely follow the overall velocity gradient
in the gas.  This shows that the more evolved pre-main sequence stars
also retain the primordial velocity structure of their parental cloud.

Observations of 3-5~Myr clusters show that the parental gas is largely
dispersed \citep[see discussion of examples
in][]{2007prpl.conf..361A}; at this time the cluster may become
unbound since the majority of the binding mass has been ejected from
the cluster \citep{ 1984ApJ...285..141L,2001MNRAS.321..699K}. The
question is whether this timescale allows for the cluster to relax.
Adopting radii of 0.5~pc for the clusters (encompassing the dense
central regions of the clusters) and using the estimated number of
stars (420, 156, 182 and 1000 stars for IC~348, NGC~1333, Serpens and
ONC), we estimate relaxation times of 2.4, 1.8, 1.8 and 6~Myr,
respectively.  The actual relaxation times will be longer since the
mass in the clusters is dominated by the gas
\citep{2001ApJ...553..744A}.  This suggests that the relaxation time
is approximately equal to or less than the gas dispersal time.  Star
formation continues in the residual cloud as the gas is being
dispersed, as shown in IC 348 and the ONC (Fig.~\ref{fig:ic348} \&
\ref{fig:onc}).  \citet{2007AJ....134..411M} estimate that the current
star formation rate in IC 348 is consistent with a constant star
formation rate over the entire lifetime of the cluster.  Thus, the
timescale for star formation, gas dispersal and relaxation are similar
\citep[cf.\ also][]{2007prpl.conf..361A}.

A picture is emerging where clusters form in filamentary, clumpy
clouds.  During the first few million years of their evolution,
clusters form stars over regions $\sim 1$~pc in diameter.  During
this time, the surrounding gas is being dispersed by winds and
radiation, eventually leading to the cessation of star formation in
3-5~Myr. Although the stars may form with a low velocity dispersion,
at least relative to local structures in the gas, collapse and
large scale velocity gradients in clouds may stir up the cluster
\citep{2007A&A...464..983P,2007arXiv0711.0391F}.  In extreme cases,
this may lead to violent relaxation, as suggested for the ONC by
\citet{2005ApJS..160..379F}. Dynamical interactions between stars and
between the stars and the cloud potential will scatter stars; however,
given that the cloud mass (which typically embodies 3/4 of the total
mass) is being ejected on a timescale similar to the relaxation time,
it is unlikely that most embedded clusters will be able to achieve
a relaxed equilibrium. One counter example may be IC 348, where the
circularly symmetric, centrally condensed configuration of this
cluster suggests that it is relaxed \citep{2007AJ....134..411M};
however, circular symmetry could also result from dynamical expansion
\citep{2005ApJ...632..397G}.   N-body simulations are needed
to fully understand what the observed cluster morphologies imply about the
dynamic state of the cluster.

\section{Local Theory of Distributed and Clustered Star Formation}

Stars form in turbulent, magnetized, clouds. The relative importance
of magnetic fields and turbulence in controlling star formation is
a matter of debate. Early quantitative models have concentrated on
the formation and evolution of individual (quiescent, low-mass)
cores and the role of magnetic fields \citep{%
1987ARA&A..25...23S,
1999osps.conf..305M}. 
More recent studies have concentrated
on the role of turbulence in cloud dynamics and core formation,
as reviewed
in, e.g., \citet{%
2004RvMP...76..125M,
2004ARA&A..42..211E,
2007prpl.conf...63B}, 
and \citet{%
2007ARA&A..45..565M}. 
Ultimately, the debate can only be settled by direct measurement
of the flux-to-mass ratio for the bulk of the molecular gas,
which is currently not available \citep{1999ApJ...520..706C}.
In the absence of
such measurements, we have to rely on indirect evidence and
theoretical arguments.

\citet{1999osps.conf...29M} argued that self-gravitating GMCs should
be magnetically supercritical by a factor of $\sim 2$ as a whole, if
turbulence and ordered magnetic fields provide comparable cloud support
\citep[e.g.,][]{2007arXiv0707.2818N}. 
Even in such a globally (moderately) supercritical
GMC, the magnetic field may still play a key role in regulating star
formation, because the formation of a typical star or small stellar
groups involves only a small sub-piece of the cloud that contains a
few to tens of solar masses. Such a region can be locally subcritical,
with a mass less than the magnetic critical mass, which can be
hundreds of solar masses for typical parameters
\citep{1999osps.conf...29M}, unless (1) a large fraction (perhaps
$\sim 1/2$ or more) of the material along a flux tube has collected in
the region,
or (2) the region is part of a substructure (i.e., a dense
clump) that is already highly supercritical to begin with.
In the former case, it would take a long time (perhaps more
than 10~Myrs) for the required mass accumulation to occur
in turbulent GMCs of tens of parsecs in size and a few
km/s in rms speed, if it happens at all. If not, locally
supercritical sub-pieces of stellar masses must be
created by some other means, most likely through
ambipolar diffusion. In the latter case, ambipolar
diffusion is not expected to play a decisive role in star
formation. The two cases may correspond to, respectively,
distributed and clustered modes of star formation \citep{%
1987ARA&A..25...23S}. 

\subsection{Ambipolar Diffusion and Distributed Star
Formation}

The best studied region of distributed star formation is the
Taurus molecular clouds. Although it is not clear whether
these clouds are magnetically supercritical or subcritical
as a whole, the dynamics of the more diffuse regions is
probably magnetically dominated.
The best evidence comes from thin strands of $^{12}$CO
emission that are aligned with the directions of the local
magnetic field (Heyer et al. 2008, preprint), where
the field is apparently
strong enough to induce a measurable difference between the
turbulent velocities along and perpendicular to the field
direction. The filamentary morphology is strikingly similar
to that observed in the nearby Riegel-Crutcher HI cloud,
mapped recently by \citet{
2006ApJ...652.1339M} 
using
21cm absorption against the strong continuum emission
towards the galactic center. Its filaments are also along
the directions of the local magnetic field, which has an
estimated strength of $\sim 30$~$\mu$G. \citet{%
1986A&A...164..328K} 
measured the line-of-sight field strength at a
nearby location, and found $B_{los}\sim 18~\mu$G, which
is consistent with the above estimate if a correction of
$\sim 2$ is applied for projection effect. The inferred
strong magnetization of this cloud may not be too
surprising, in view of the result that cold neutral
HI structures are strongly magnetically subcritical
in general \citep{%
2005ApJ...624..773H}. 
If diffuse
molecular clouds such as the Taurus clouds are formed
out of such HI gas, it is not difficult to imagine that
at least some of them will be subcritical as well.
If the bulk of a molecular cloud is indeed magnetically
subcritical, then the well-known low efficiency of star
formation can be naturally explained: the formation
of dense, star-forming, magnetically supercritical cores
is regulated by ambipolar diffusion, which is generally
a slow process, with a time scale an order of magnitude
(or more) longer than the local free fall time
\citep{%
1987ARA&A..25...23S,
1999osps.conf..305M}. 

For relatively diffuse molecular gas with an $A_V$ of
a few, the problem may be the opposite: the ambipolar
diffusion time scale is too long to allow for significant
star formation in a reasonable time due to ionization
from UV background \citep{%
1989ApJ...345..782M,
1995ApJ...442..186M}. 
In order to form stars in a reasonable time scale, the
rate of ambipolar diffusion must be enhanced \citep{%
2002ApJ...567..962Z,
2002ApJ...570..210F}. 
This is where turbulence can help greatly.
Supersonic turbulence naturally
creates dense regions where the background UV photons
are shielded, and where the gradient in magnetic field
is large; thereby accelerating ambipolar diffusion.
The strong magnetic field, on the other hand, prevents
too large a fraction of the cloud mass being converted
into stars in a turbulence crossing time. In this
hybrid scenario of distributed star formation in relatively
diffuse molecular clouds, both magnetic fields and
turbulence play crucial roles: the star formation is
accelerated by turbulent compression but regulated by
magnetic fields through ambipolar diffusion
\citep{%
2004ApJ...609L..83L}. 

The above scenario is illustrated in \citet{%
2005ApJ...631..411N}, 
using 2D simulations of a sheet-like, magnetically
subcritical cloud (with an initial dimensionless
flux-to-mass ratio $\Gamma=1.2$ everywhere). It is stirred
with a (compressive) supersonic turbulence of rms Mach
number ${\mathcal M}=10$ at $t=0$. The strong initial
magnetic field prevents the bulk of the strongly shocked
material from collapsing promptly. The shocked material
is altered permanently, however; its flux-to-mass ratio
$\Gamma$ is reduced through enhanced ambipolar diffusion, which
produces filamentary supercritical structures that are
generally long lived. Only the densest
parts of the filaments---the dense cores---are directly
involved in star formation. Even the dense cores are
significantly magnetized, with flux-to-mass ratios about
half the critical values or more. The strong magnetization
ensures that ambipolar diffusion continues to play a
role in the core evolution, reducing their internal
velocity dispersions to typically subsonic levels, as
observed \citep{%
1995mcsf.conf...47M}. 
Since only a fraction of the cloud
material is magnetically supercritical (and thus capable
of forming stars in the first place), and only a small
fraction of the supercritical material is directly
involved in star formation, the efficiency of star formation
is naturally low.

An attractive feature of the ``turbulence-accelerated,
magnetically regulated'' star formation in diffuse,
subcritical clouds (or sub-regions) is that the stars
are expected to form at distributed locations. This is
because the self-gravity, although important on the
scale of individual supercritical cores, is canceled
out to a large extent, if not completely, by magnetic
forces on large scales. The cancelation makes the
clustering of dense cores difficult, unless the cores
happen to be created close together by converging flows.
By the same token, star clusters are more likely formed
in dense, self-gravitating clumps that are magnetically
supercritical as a whole, as we discuss below.

Even though the principle of the ``turbulence-accelerated,
magnetically regulated'' star formation is straightforward, and well
illustrated by 2D simulations, much work remains to be done to firm it
up. Among the needed refinements is the extension of the 2D
calculations to 3D. A first step in this direction has been taken by
\citet{ 2007MNRAS.380..499K} 
and Nakamura \& Li (in preparation).  Another
improvement would be in the maintenance of turbulence, which will be
more difficult, given our limited understanding of the origins of
turbulence in molecular clouds in general, and in diffuse regions of
distributed star formation in particular.

A region of inefficient, distributed star formation of
considerable current interest is the Pipe nebula.
\citet{%
2007A&A...462L..17A} 
found more than 150 dense quiescent cores
with a mass distribution that resembles the stellar IMF.
The cores are distributed along a filament of more than
10~pc in length. None of the cores except the most massive
one has yet collapsed and formed stars. This is difficult
to understand, given their rather short free fall times,
unless the core creation is well synchronized. The
synchronization is most naturally done by a large
scale shock. In such a case, the strong compression is
expected to induce fast motions inside the filaments and
cores, unless the cores are magnetically cushioned. The
Pipe nebula may turn out to be a good example of
compression-induced, magnetically regulated star formation.

\subsection{Turbulence, Gravity, and Cluster Formation}

The most detailed simulations of cluster formation are performed using
the SPH technique, which generally does not include magnetic fields
\citep[see, however,][]{%
2007MNRAS.377...77P}. 
These simulations are well suited for
studying the interaction between turbulence compression, gravitational
collapse, and especially mass accretion onto collapsed objects
\citep[see][for a recent review]{%
2007prpl.conf..149B}. 
They fall into two
categories, with either a freely decaying
\citep[e.g.,][]{%
2003MNRAS.339..577B,
2003MNRAS.343..413B} 
or a constantly driven turbulence
\citep{%
2001ApJ...550L..77K}. 
In the former, the initial turbulence typically generates
several dense clumps, in which small stellar groups
\citep{%
2003MNRAS.339..577B} 
or sub-clusters
\citep{%
2003MNRAS.343..413B} 
form, depending on the
number of Jeans masses initially contained in the cloud. The most
massive member of each sub-cluster typically sits near the bottom of
the gravitational potential well of the local collection of gas and
stars, gaining mass preferentially through competitive accretion.
The sub-clusters of stars and gas merger together at later times,
making deeper gravitational potentials where the more massive stars
near the bottom of the potential well can accrete preferentially,
growing to higher masses.

The most attractive feature of the non-magnetic SPH calculations of
cluster formation is that an IMF that broadly resembles the observed
one is produced. In particular, the characteristic mass near
the knee of the IMF is attributed to the initial cloud Jeans mass
\citep{%
2006MNRAS.368.1296B}, 
and the steeper, Salpeter-like mass
distribution above the knee is mostly shaped by competitive
accretion. A potential difficulty is that the Jeans mass (and thus the
characteristic mass of IMF) is expected to vary from region to region
because of variations in both density and temperature. There is little
observational evidence to support this expectation. The difficulty can
be alleviated to some extent by the heating and cooling behaviors of
molecular gas, which tend to set a characteristic density below
(above) which the temperature decreases (increases) with density
\citep{%
2005A&A...435..611J,
2006MNRAS.368.1296B}. 
Whether the
thermal regulation of Jeans mass works for very dense clusters where
the average Jeans mass is expected to be very small remains to be seen
\citep{%
2007ARA&A..45..565M}. 

A problem with cluster formation in decaying turbulence is that star
formation may be too rapid. According to
\citep{%
2001ApJ...550L..77K} 
$\sim 20-30\%$ of the cloud mass is accreted by sink particles in one free
fall time $t_{ff}$. The value of $t_{ff}$ depends on the density (and thus
the initial Jeans mass $M_J$ for 10~K gas).  For $M_J=1~M_\odot$
(needed to reproduce the knee of IMF), $t_{ff} \approx
10^{5}$~yrs. This may be short compared to the lifetimes (typically 1
Myrs or more) estimated for nearby embedded clusters, particularly
those that include a substantial population of relatively evolved
Class II and Class III objects. Increasing the free-fall time to 1
Myrs would increase the Jeans mass to $10~M_\odot$, which would
produce too high a mass for the knee of the IMF, unless the conversion
of dense, self-gravitating gas into individual stars is inefficient
\citep{%
1999osps.conf..193S,
2000ApJ...545..364M}. 
Alternatively, the rate of
star formation can be reduced significantly, if the turbulence is
constantly driven on a small enough scale
\citep{%
2001ApJ...550L..77K}. 
The small-scale driving may, however, modify the hierarchical cluster
formation observed in the simulations of the decaying turbulence case,
and affect competitive accretion and thus the high mass end of the
IMF. Indeed, \citet{%
2001ApJ...550L..77K} 
and \citet{%
2003ApJ...585L.131V} 
suggested that turbulence driven on small scales may correspond to
distributed star formation, whereas that on large scales to clustered
star formation.  According to these works, large-scale driving does not
slow down star formation significantly, so the problem of too rapid star
formation in clusters may remain (but see Sect.\ \ref{global} below).
The problem can be alleviated somewhat by the
inclusion of a strong magnetic field
\citep[e.g.,][]{%
2001ApJ...547..280H,
2005ApJ...630L..49V,
2007MNRAS.tmp..930T}. 
In the next
subsection, we examine the role of protostellar outflows in slowing
star formation by replenishing the turbulence in localized regions
of active cluster formation.

\subsection{Cluster Formation in Protostellar Turbulence}

The possibility of outflows replenishing the energy and momentum
dissipated in a star-forming cloud was first examined in detail by
\citet{%
1980ApJ...238..158N}. 
They envisioned the star-forming clouds to be constantly
stirred up by the winds of optically revealed T Tauri stars. The idea
was strengthened by the discovery of molecular outflows
\citep{%
1989ApJ...345..782M}, 
which point to even more powerful outflows from the stellar vicinity
during the embedded, {\it protostellar} phase of star formation
\citep{%
1985ARA&A..23..267L,
1996A&A...311..858B}. 
\citet{%
1999osps.conf..193S} 
estimated the momentum output from protostellar outflows
based on the Galactic star formation rate, and concluded that it is
sufficient to sustain a level of turbulence of $\sim 1 - 2$ km/s,
similar to the line widths observed in typical GMCs. If the majority
of stars are formed in localized parsec-scale dense clumps that occupy
a small fraction of the GMC volume
\citep{%
1991ApJ...368..432L}, 
their ability to influence the dynamics of the
bulk of the GMC material will probably be reduced; other means of
turbulence maintenance may be needed in regions of relatively little
star formation, as concluded by
\citet{2005AJ....129.2308W} 
in the case of the Perseus molecular cloud. The concentration of star
formation should, however, make the outflows more important in the
spatially limited, but arguably the most interesting regions of a
GMC---the regions of cluster formation, where the majority of stars
are thought to form.

The importance of outflows on cluster formation can be illustrated
using a simple estimate. Let the masses of the cluster-forming dense
clump and the stars formed in it be $M_c$ and $M_*$, respectively. If
the outflow momentum per unit stellar mass is $P_*$, then there is
enough momentum to move all of the clump mass to a speed
\begin{equation}
v\sim {M_* P_* \over M_c} \sim \epsilon P_* =5~km/s \left({\epsilon\over
0.1}\right) \left({P_*\over 50~km/s}\right),
\label{vel}
\end{equation}
where $\epsilon=M_*/M_c$ is the star formation efficiency (SFE).
For embedded clusters, the SFE can be $10\%$ or more
\citep{%
2003ARA&A..41...57L}. 
The value of $P_*$ is somewhat uncertain. For low
mass stars,
\citet{%
2007ApJ...662..395N} 
estimated a plausible range
between $10 - 100$~km/s. For the best studied CO outflow-driving
early B stars, the value is estimated at $50-100$~km/s
(D. Shepherd, priv.\ comm.). For the fiducial values of $\epsilon$
and $P_*$ adopted in equation~(\ref{vel}), the
outflow-driven average velocity is
comfortably above the typical turbulence velocity of 1-2~km/s
observed in nearby cluster-forming clumps. Indeed, if all of
the cluster members were to form simultaneously, the clump would
quickly become unbound. If the stars are formed more gradually
(as evidenced by the presence of objects in a wide range of
evolutionary states, from prestellar cores to Class III sources),
then there is the possibility for the outflows to replenish the
dissipated turbulence, keeping the cluster formation going for
a time longer than the global free fall time.

\begin{figure}[t]
\includegraphics[width=10cm,angle=0]{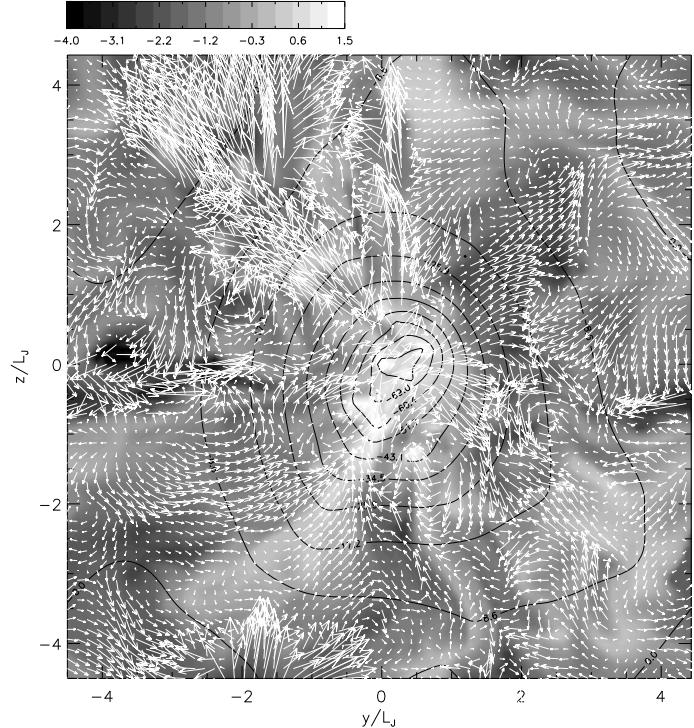}
\caption{Snapshots of a 3D MHD simulations of cluster formation
including feedback from protostellar outflows. Plotted are the
volume density map, velocity vectors, and contours of gravitational
potential on a slice through the simulation box. For details,
see \citet{%
2007ApJ...662..395N}. 
}
\label{outflow}
\end{figure}

Quasi-equilibrium cluster formation in outflow-driven turbulence
is illustrated in Fig.~\ref{outflow}. It is clear that dense
cores tend to collect near the bottom of the gravitational potential
well, where most of the stars form. Most of the outflow momentum
is coupled, however, into the envelope, where most of the clump
mass resides. The coupling of momentum into the envelope is
facilitated by outflow collimation, which
enables the outflows to propagate further away from the central
region, and drive motions on a larger scale that would decay
more slowly. Gravity also plays an important role in
generating the turbulent motions by pulling the slowed down
material towards the center and setting up a mass circulation
between the core and envelope. For parameters appropriate for
nearby embedded clusters such as NGC 1333 ($\sim 10^3$~M$_\odot$
in mass and $\sim 1$~pc in size), the gravitationally induced
infall and outflow-driven expansion are roughly balanced,
creating a quasi-static environment in which the clump
material is converted into stars at a relatively leisure
pace. For the particular example shown, the average star
formation efficiency per global free fall time is about
$3\%$. This value is in the range inferred for NGC 1333
\citep{%
2007ApJ...662..395N} 
and other objects
\citep{%
2007ApJ...654..304K}. 

The protostellar turbulence in regions of active cluster formation
is a special type of driven turbulence. It is driven
anisotropically over a range of scales, as the collimated outflow
propagates progressively further away from the driving source
\citep{%
2006ApJ...646.1059C,
2006ApJ...649..845S,
2007ApJ...668.1028B}. 
There is a break in the energy spectrum near the characteristic
length scale of the outflows, as predicted analytically by
\citet{%
2007ApJ...659.1394M}. 
The break may provide a way to distinguish the protostellar
turbulence from other types of turbulence.

It is interesting to speculate how massive stars may form in
protostellar outflow-driven turbulence. One possibility, pointed out in
\citet{%
2006ApJ...640L.187L}, 
is that, when the central region of the
cluster-forming clump becomes dense enough, the outflows from the low
mass stars may become trapped. The trapping of outflows reduces the
stellar feedback into the massive envelope, which leads to more mass
falling to the central region, which in turn makes it more difficult
for the outflow to get out. It may lead to a run-away collapse of the
central region, with a large mass infall rate that is conducive to
massive star formation \citep{%
2002Natur.416...59M}. 
The massive stars formed in
this scenario will naturally be located near the bottom of the
gravitational potential well, for which there is now growing
observational evidence
\citep{%
2005IAUS..227...86G,
2005Ap&SS.295....5C}. 
The trapping of
outflows as a cluster-forming clump condenses may also provide a
natural explanation for the accelerating star formation inferred for
the Orion Nebula cluster
\citep{%
2006ApJ...644..355H} 
and others
\citep{%
2000ApJ...540..255P}. 
Detailed numerical simulations are needed to firm up
this supposition.

\section{Global Theory of Star Formation in Turbulent Clouds}
\label{global}

\def\sigmaV{\sigma_{v,3D}} \def\rhoL{\rho_{L}} \def\pV{p_v}
\def\pN{p_n} \def\vrms{v_{\mbox{rms}}}

Giant molecular clouds and their smaller scale constituents, star forming clumps and
prestellar cores, obey several remarkable scaling relations, which indicate that
star formation to a significant extent exhibits a generic behavior.  The most
important of these scaling relations are:

\begin{enumerate}

\item
Larson's velocity-size relation \citep{%
1979MNRAS.186..479L}, 
which expresses that the (3D) rms velocity dispersion $\sigmaV$ of structures of
size $L$ scales as $\sigmaV \sim L^{\pV}$, where $\pV \approx 0.4$ \citep%
[][]{%
1981MNRAS.194..809L}. 
Recently \citet{%
2007IAUS..237....9H} 
have shown that the relation is very tight when measured in a consistent manner over a
limited range of scales, and that {\em the normalization is essentially the same} in
violently star-forming and nearly quiescent molecular clouds
\citep{%
2006ApJ...643..956H}. 

\item
Larson's velocity-mass relations \citep{%
1981MNRAS.194..809L}, 
which expresses that the densest structures on each scale are in near-virial equilibrium.

\item
The cluster mass function, which is a power law
with slope of $d N/d M$) slightly less steep that -2 \cite[][and references therein]{%
2006astro.ph.11586D}
---so stellar birth is distributed
nearly evenly over mass, although a slightly larger number of stars are
born in rich clusters.

\item
The Core Mass Function (CMF) and the Initial Mass Function (IMF), which are approximate
power laws for large enough masses.  Individual examples may show
deviations, due to statistics or other influences, but the general power law
trends are well established.

\end{enumerate}
These structures thus appear to be members of a hierarchy of
scales, characterized by approximate scaling laws. Two additional, and crucially
important, properties of this hierarchy are that:

{
\begin{enumerate}
\renewcommand{\labelenumi}{(\alph{enumi})}
\item
On each scale the actual lifetimes of the structures appear to be
longer than their estimated turbulent decay times (as given
by the turbulence crossing times). To maintain their
\citep[rather precise; cf.][]{%
2006ApJ...643..956H} 
velocity normalization in the hierarchy, the motions on each scale thus
need significant driving.

\item
A related but largely independent property is that the mass drain due to the
star formation process is approximately constant across scales, at
a level of only a few percent per unit free fall time
\citep[cf.][and references therein]{%
2007ApJ...654..304K}.

\end{enumerate}
}

The various properties enumerated above are linked in ways that are
likely not to be coincidences: items (a) and (b) above imply that the
draining of mass from each scale per unit crossing time is quite
small.  This property is in principle independent of the property that
the structures appear to be long lived; one could imagine the
structures to have a small draining rate but still be quite
short-lived.  The opposite could also be the case; the structures
could be drained more rapidly by star formation but could still be
long lived if the geometrical shapes of structures were somehow
maintained as they were being drained.

\subsection{Questions and Difficulties}

Most of these properties are difficult to explain: the small rate of star formation per free
fall time
\citep{2007ApJ...654..304K}, 
the driving source of the motions, the regularities of the cluster mass function, the core mass
function, and the stellar initial mass function are all long standing problems.

One can attempt to ``explain'' Larson's density and velocity relations as a
consequence of ``virialization'', since they are roughly consistent with virial
equilibrium. With velocity scaling as $L^{-1/2}$ (rather than the observed
exponent $\pV \approx 0.4$) and density as $L^{-1}$, one could imagine that gravity
maintains the velocity spectrum, as structures shrink in size and descend into
ever deeper potential wells.
That, however, still begs the question ``why would that happen''; specifically,
why would this happen in such a way that the life times remain long and the
draining rates at each level in the hierarchy remain nearly constant (and low)?
One would rather expect that the characteristic time scale of such a cascade would
be similar to the local gravitational free fall time. The driving-by-gravity
explanation is also very difficult
to reconcile with the fact that potential wells actually get {\em shallower}
with decreasing size, rather than deeper!

Likewise, it is very difficult to explain how local feedback could,
without exceptional fine tuning, both maintain a power law scaling of
the velocity with size---the energy input rate would have to match
the turbulent dissipation rate and its variation with scale over a
large range of scales---and maintain the proper
amount of support against gravitational collapse to keep the draining
rate constant and low over a range of structure
sizes.

One is lead to conjecture then that local feedback, rather than {\em maintaining}
the observed scaling laws over a range of scales instead may be responsible for
{\em breaking} those scaling laws at local sites with rapid star formation; as
observed in for example CS clumps \citep{%
2007ApJ...654..304K}, 
where the mass density (and possibly also draining rates) exceed values
expected from the scaling relations. But that still
leaves the questions of explaining the power law scalings that hold at most
scales and locations, of explaining the maintenance of these motions against
dissipation, and of explaining the constancy of draining rates over all other
scales.

In fact, most of the scaling and lifetime properties may be explained,
in a consistent and complete manner, by essentially ``turning the
explanation up-side-down''; it is not the gravity that is in the lead
but the kinetic energy, cascading from large to small scales, with
gravity picking up those exceptional regions of space where the local
gravitational potential energy dominates the local kinetic energy
(i.e.  the local virial parameter, as given in
Eq.~\ref{equation}, happens to be small---cf.\ Fig.\
\ref{alpha}).

\subsection{A Turbulent Cascade Origin of the Scaling}

A suitable entry point for discussion of the various scaling
properties is Larson's velocity-size relation.  As shown by \citet{%
2002ApJ...569..841B} 
and \citet{%
2002ApJ...573..678B} 
such a scaling follows inevitably from the scaling relations of
supersonic turbulence. A number of observationally well defined
diagnostics are in agreement with the corresponding diagnostics
extracted from simulations of supersonic (and super-Alfv{\'e}nic)
turbulence \citep{%
1997ApJ...474..730P,
1998ApJ...504..300P,
2001ApJ...553..227P,
2004ApJ...604L..49P}. 
One can even measure, using the method of \cite{%
2000ApJ...537..720L} 
the velocity power spectrum exponent of the turbulence; the result is a
wavenumber exponent $\approx -1.8$, consistent with a velocity-size
scaling with an exponent $\pV \approx 0.4$ \citep{%
2006ApJ...653L.125P}.

So, what we are observing as Larson's velocity-size scaling law is,
most likely, just the inevitable and robust consequence of a cascade of
kinetic energy from the largest (injection) scales, which occur at scales
of the order of the galactic disk thickness, where kinetic energy
may be contributed by several sources \citep{%
2005A&A...436..585D,
2007ApJ...665L..35D,
2006ApJ...649L..13K,
2007IAUS..237...70O,
2007ARA&A..45..565M}. 
No matter how that energy is fed into the medium it must cascade to very
small scales before it is dissipated.  This occurs, in supersonic turbulence
as well as in subsonic turbulence, via a turbulent cascade across an
``inertial range'', where the denomination ``inertial'' signifies that
the motions are maintained by inertia.

This explains both the power law distribution of velocity (and hence Larson's
velocity-size relation), and the source of apparent ``driving'' over a range
of scales.  As already shown by \citet[cf.\ his Fig.\ 1]{%
1979MNRAS.186..479L} 
the power law scaling continues to sizes comparable to the thickness of the galactic disk,
where it exhibits a break (at larger scales the velocity dispersion is
mainly associated with fluctuations in the galactic disk rotation rate).
The title of Larson's (1981) paper was indeed, and very appropriately, ``Turbulence and
Star Formation in Molecular Clouds''!

The super-Alfv{\'e}nic turbulence scenario may also be used to derive the form
of the stellar IMF analytically \citep{%
2002ApJ...576..870P} 
and to explain how brown dwarfs can form by the same mechanism as normal stars
\citep{%
2004ApJ...617..559P}. 

\subsection{Interaction of Turbulence and Selfgravity}

One might object to the comparisons of diagnostics from simulations of
supersonic turbulence without selfgravity with observations of molecular
clouds since these, according to the previous discussion, are often close to
virial equilibrium, and it would thus seem that selfgravity cannot be
ignored.

One could, on the other hand, also argue that the success of such exercises
might indicate that selfgravity is only important in local regions, and that
these are sufficiently small not to disturb the comparisons significantly.
Such a conjecture can actually be confirmed from an analysis of locally
evaluated virial ratios (`virial parameters'), which we may define as
\citep[cf.][]{%
2005ApJ...630..250K}
\begin{equation}\label{alpha.eq}
\alpha = {5 \vrms^2\over 6 G \rho L^2} ,
\label{equation}
\end{equation}
where $\vrms$ is a local (3-D) rms velocity dispersion, measured in some
neighborhood of size $L$ around comoving, lagrangian tracer points where the
local average density is $\rho$.

\begin{figure}
\PSFIGG{file=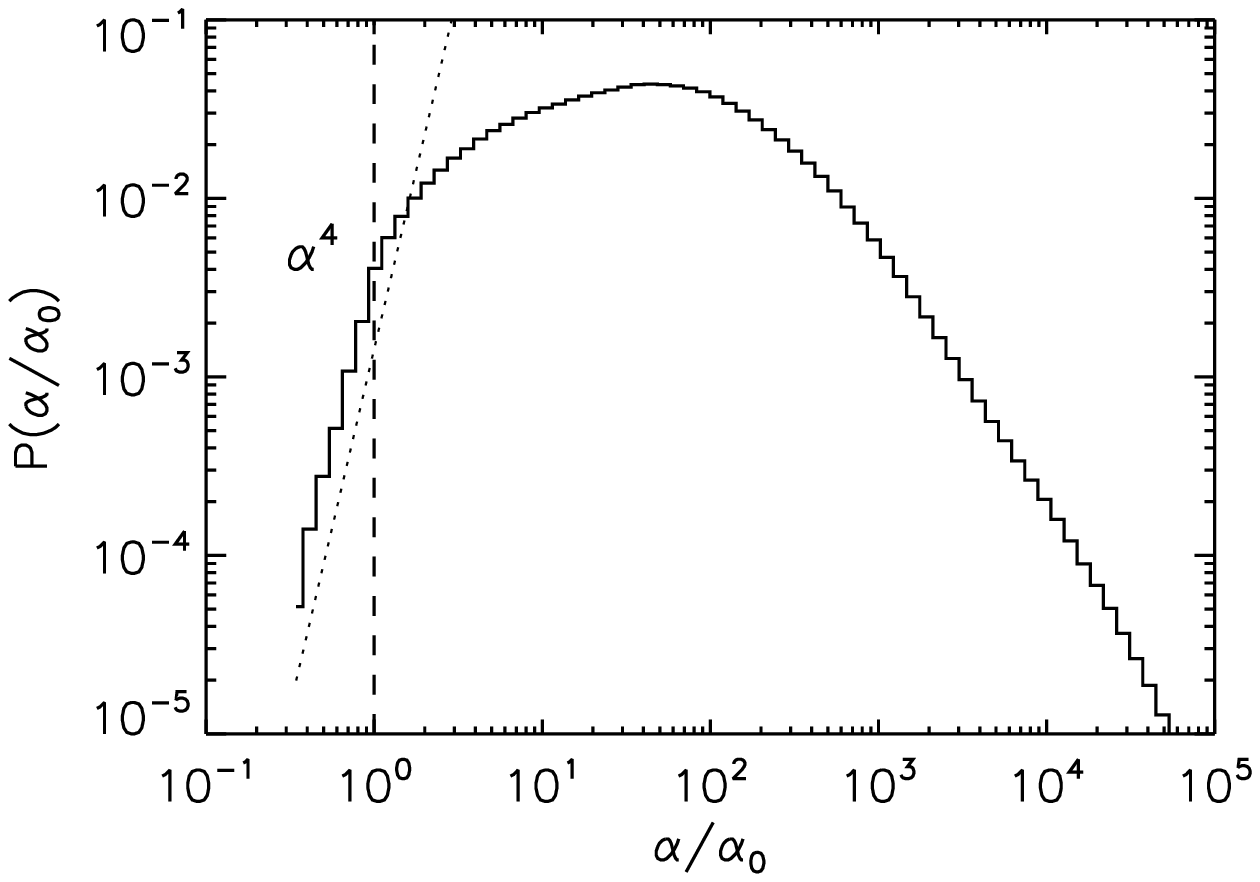,width=2.5in}{file=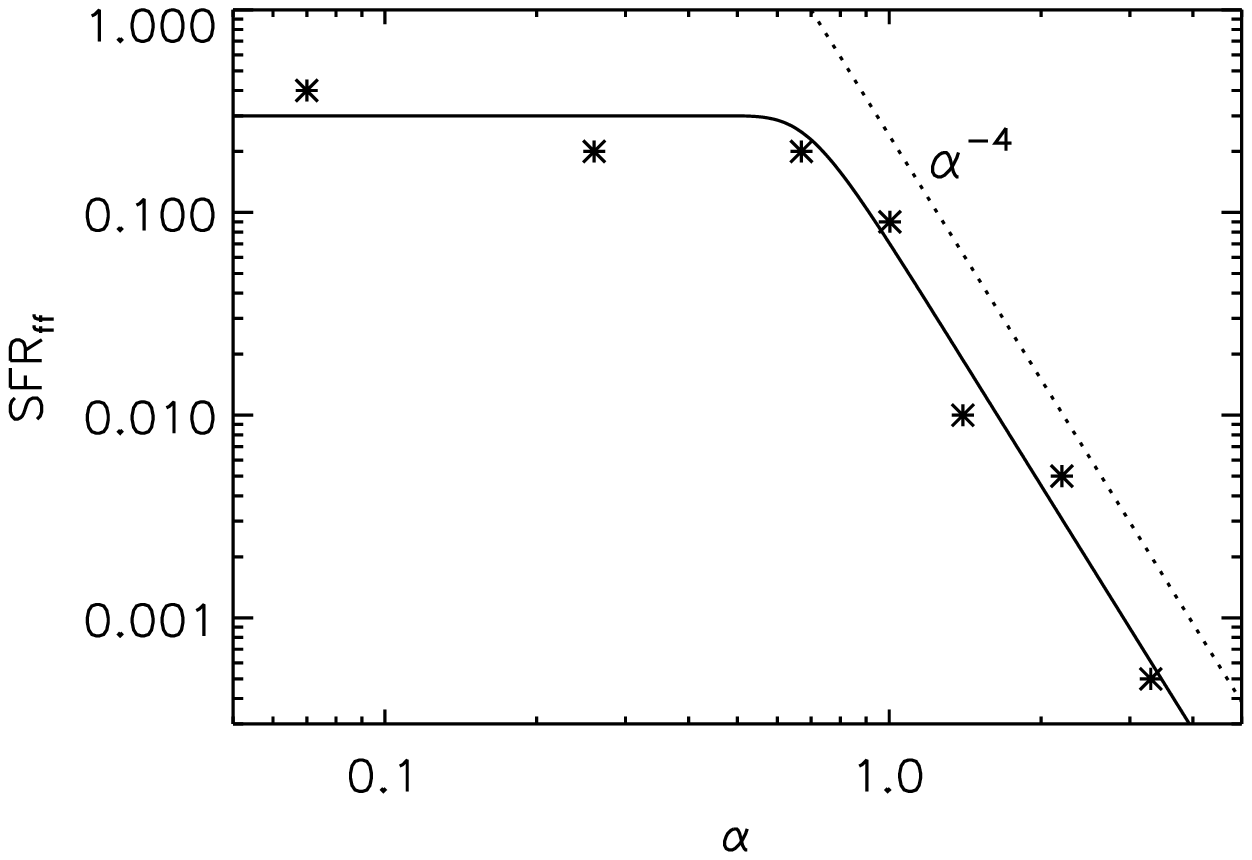,width=2.5in}
\caption{The mass weighted probability distribution of $\alpha$ (left),
normalized to the virial ratio for the whole domain, in a numerical simulation
of driven Mach=18 supersonic turbulence with selfgravity (Padoan \& Nordlund
in prep.; resolution $1000^3$), and the dependence of the star formation rate on
the large scale $\alpha$ (right) for a number of such simulations (stars---%
the broken power law curve is only suggestive).}
\label{alpha}
\end{figure}

The left hand side panel in Fig.\ \ref{alpha} shows an example of the mass
weighted probability distribution of the local virial ratio, using values of
$\vrms^2$ and $\rho$ evaluated locally.  It is obvious that a large fraction
of the mass has local values of $\alpha$ much larger than the virial ratio
$\alpha_0$ based on the rms velocity and average density for the whole
domain, and that only a small fraction of the mass ($\sim$1\% in this case)
has a smaller value.

A consideration of how local shearing motions are modified by compression
explains this behavior.  Suppose we have
identified a region of size $L$ with a small density enhancement (and thus
potentially a selfgravitating `object'), which initially has a value of
$\alpha$ close to the value for the whole box and a density close to the mean
density.  If we choose $L$ to be for example $1/4$ of the size of the full box
then $v$ must likewise be $1/4$ of the average rms value to obtain the same
$\alpha$, so the initial selection is already limited to a subset of the volume
(in the rest of the volume a local $\alpha$ evaluated in this fashion would be
larger, even initially).

In a local neighborhood, and for a limited time, one can assume the local,
comoving dynamics to be subsonic.  As can be confirmed by data-browsing, the velocity
field in supersonic turbulence consists of patches with smooth flow, separated
(ideally) by discontinuities of the velocity and the mass density.  With smooth
velocity across a patch one can subtract off the mean speed and consider the
local velocity field, which can be split into a soleniodal and a compressive
part.  In a sufficiently small neighborhood in space the flow remains subsonic
with respect to the mean, and in a sufficiently small neighborhood in
time---until the patch hits a shock---one may thus consider the local dynamics
to be subsonic, smooth and continuous.

One consequence of this is that, within a thus limited neighborhood in space
and time it makes sense to think in terms of approximate conservation of
local angular momentum.  The subsonic nature of the local dynamics means,
for example, that convergent motions couple to shear and rotations
in the well known sense that contraction tends to cause spin-up.

Now suppose the region in question is compressed to a tenth of its linear size,
with compression taking place in two spatial directions.
The cross section decreases with a factor of 100, but the density
increases correspondingly, so the denominator in Eq.\ \ref{alpha.eq} is
unchanged. Because of the tendency to locally conserve angular momentum the
$v^2$ in the numerator tends to increase by a large factor, and one finds
that the $\alpha$ characterizing this local region has grown tremendously,
which prevents the object from becoming selfgravitating.

In general one finds that in regions of compressions, which are
otherwise the locations that are most favorable for creating large mass
densities, the tendency for growth of vorticity more than counters the
growing influence of selfgravity, which makes it hard to achieve collapse
under selfgravity, even when the virial parameter measured on the large
scale is of the order unity.

There is thus a purely dynamical mechanism available to explain why
structures on all scales in GMCs tend to survive longer than would be
indicated by estimates based on the large scale virial parameter
(such an estimate is exactly what is being used when one says ``it is remarkable
that these structures live longer than a few free fall times'').  The
mechanism is even {\em unavoidable}; there is no way that this could {\em not}
happen, in cases where the virial parameter based on the large scale is
larger than unity!

When, on the other hand, the large scale virial parameter is less than
unity opportunities rapidly open up for local collapse; the fraction of
compressive, converging flows that already initially have low $\alpha$
increases, and a rapidly growing number of compressed regions find
themselves to have local virial ratios low enough for collapse.  To a
first approximation one can see the dependence of the draining rate on
$\alpha$ (right hand side panel of Fig.\ \ref{alpha}) as resulting from
a shift of the PDF of $\alpha$ (left hand side panel of Fig.\ \ref{alpha},
considered as a function $\alpha$ rather than $\alpha/\alpha_0$).

Such an argument may be applied recursively.  Structures that are much
smaller than the initial size but that still have low enough local virial
ratios for collapse can be taken as a basis for `renormalization'.
One concludes that, for any large scale virial ratio, a hierarchy of
smaller structures exists, with local virial ratios of the order unity,
and with internal structures consisting of yet smaller structures
with local virial ratios of the order of unity.  The hierarchy must
be such as to conserve the draining rate at each scale; since the
smaller scale free fall times are shorter than the larger scale free
fall times the smaller scales are ``slaved'' to the larger scales.

\subsection{Selfregulated Star Formation}
One very important aspect of the difficulty to achieve locally small
virial ratios when the large scale virial ratio is larger than unity is
that the rate of gravitational collapse of mass out of this hierarchy
(which presumably is a good proxy of the star formation rate---here
denoted SFR), turns out to be extremely sensitive to the large scale
virial ratio.

The right hand side panel in Fig.\ \ref{alpha} illustrates the point, by
showing a summary of the star formation rates achieved in a number of
simulations where the large scale alpha was varied, both by changing the mean
density and by changing the Mach number (Padoan \& Nordlund, in prep.). The
functional behavior is such that the SFR is approximately proportional to
$\alpha^{-4}$ for $\alpha$ larger than unity, while for smaller alpha the
SFR is roughly constant and very large ($\sim 30$ \% per free fall
time).

\citet{%
2005ApJ...630..250K} 
arrived at an analytical estimate of the dependence of the SFR on
$\alpha$ that is somewhat less steep but qualitatively similar,
based on results from numerical simulations by  \citet{%
2001ApJ...550L..77K} 
and \citet{%
2003ApJ...585L.131V,
2005ApJ...630L..49V}.

In the regime where the SFR is nearly constant it is significantly
larger than values characterizing star formation in the Milky Way \citep{%
2007ApJ...654..304K}. 
We thus conclude that galactic star formation is operating in a regime
where $\alpha$ is on the steep, power-law part of the dependence.

Does that then not require an unlikely fine tuning? Why would the large scale
$\alpha$ happen to have a value such that the SFR agrees with the observed
one?  Well, suppose that it did not:  if the value of $\alpha$
were too large there would be practically no star formation, the driving of
the ISM from massive stars and SNe would cease, the level of turbulence would
be reduced, the scale height of the ISM would drop and the large scale $\alpha$
would decrease---both because of reduced velocities and because of increased
densities.

Conversely, if $\alpha$ were too small the SFR would be too high (relative
to the level required to sustain the ISM turbulence), and the situation
would rectify itself via the same feedback mechanism.

We conclude that the mechanism that determines the level of the
star formation rate is large scale stellar feedback.  Note also that this
only works if the driving of the interstellar medium turbulence is, to
a significant but not necessarily exclusive extent, due to large scale
feedback from star formation.  That this is so is already abundantly
established by elaborate numerical simulations, in particular the ones
by \citet{
2005A&A...436..585D,
2007ApJ...665L..35D}. 
Such a scenario is also entirely consistent with the threshold assumptions
about star formation generally made in models of galaxy formation
\citep[e.g.\ ][]{%
2007ApJ...668..826C,
2007ApJ...668....1N} 
and with the Kennicut-Schmidt law \citep{%
1998ApJ...498..541K,
2007IAUS..237..311K}. 

\section{Concluding Remarks}

The last decade has brought significant observational and theoretical
progress to the study of star formation in molecular clouds,
particularly in observations of embedded clusters and simulations of
turbulent clouds. With large scale surveys of both molecular clouds
and their populations of protostars and pre-main sequence stars now in
hand, and with increasingly sophisticated numerical models, there is
excellent potential to rapidly advance our understanding of the
macrophysics of star formation.

To do so will require better means for comparing observations and
simulations.  Observers and modelers need to devise a set of common
``observable'' statistics that may be used to perform quantitative
comparisons.  Given the advances in measuring the distribution,
evolutionary state, and even kinematics of YSOs and dense cores, it is
of great interest to relate the properties of the sink particles in
simulations of star formation to the observed properties of dense
cores and YSOs.  In particular, it is important to determine whether
the observed spatial distributions of protostars and more evolved
pre-main sequence stars can be reproduced. The evolution of clusters
should also be explored through N-body simulations with realistic and
time variable gas potentials to better understand the dynamical
states implied by their observed morphologies.

There are areas where significant progress is needed on the
observational front. Much progress is still needed to understand the
ages and age spreads of embedded populations of YSO. These are an
important constraint on the lifetimes of clouds and the duration of
star formation. Significant advances have been made in the
measurements of magnetic fields; however, since these measurements
often target regions of active star formation, the field structure in
the bulk of cloud material is poorly constrained. Large-scale mapping
of the strength and direction of the magnetic field is needed to
ascertain its dynamical role, particularly in relatively diffuse
regions of clouds, where the field may play a dominant role.

Theoretically, there is a need to tie local simulations of star
formation to global simulations of structure formation on GMC scales
and beyond. On the GMC and smaller scales there are additional effects
not included in this review.  These include the generation of small
scale structure from global cloud collapse
\citep{2004ApJ...616..288B}, generation of
structure by thermal instabilities at the interfaces between the warm
ISM and molecular clouds
\nocite{2007arXiv0709.2451H} (Heitsch et al., 2007),
self-shielding and
the transition from atomic to molecular phase in the interior of
molecular clouds.  In order to avoid the need for ad hoc initial and
boundary conditions such modeling may be performed as sub-sets of
larger scale and longer duration global simulations, as often done in
the study of galaxy formation in the context of cosmological structure
formation simulations.

On larger scales, simulations need to account for the life cycle
of molecular gas, and determine whether ``lifetime'' is even a meaningful
concept in the context of molecular clouds.  Relevant questions to
investigate include ``Given a snapshot of a molecular cloud, how long
time does it take before half of the molecular gas present in the
snapshot is recycled?'', ``How much of that went into stars?'',
``How much was turned into warm and hot ISM, respectively'', ``How
much new molecular gas was added in the mean time?'', ``How much of
the feedback is due to winds, UV- radiation, and supernovae and on what
scales is this feedback deposited?''.  Such questions
may be answered by realistic high resolution global
ISM simulations \citep[such as the ones by][]{2007ApJ...665L..35D},
which include the vertical structure of the galactic disk, realistic
cooling functions and UV heating, selfgravity, and magnetic fields.
A next step in such simulations could be to let the star formation rate
be determined self-consistently, by including approximations of how
the star formation rate depends on the virial ratios of star forming
structures and their average magnetic fields.

{\bf Acknowledgements:} {\AA}N acknowledges support from the Danish
Natural Science Research Council and the Danish Center for Scientific
Computing.  The work of ZYL is supported in part by NASA (NNG05GJ49G)
and NSF (AST-0307368) grants.  This work is based in part on
observations made with the Spitzer Space Telescope, which is operated
by the Jet Propulsion Laboratory, California Institute of Technology
under a contract with NASA. Support for the work of STM was provided
by NASA through an award issued by JPL/Caltech. All authors
participated in the KITP program ``Star Formation Through Cosmic
Time'', supported in part by the National Science Foundation under
Grant No. PHY05-51164.


\input{abbrev}

\bibliography{ADS}


\end{document}

%% file: abbrev.tex
\def\newblock{}
\def\aap{AA}
\def\aj{AJ}
\def\apj{ApJ}
\def\apjl{ApJL}
\def\apjs{ApJS}
\def\baas{BAAS}
\def\nat{Nature}
\def\mnras{MNRAS}
\def\apss{APSS}
\def\araa{ARAA}
\def\angst{\AA}